\documentclass[prl,showpacs,twocolumn,superscriptaddress,floatfix,amsmath,amssymb]{revtex4}

\usepackage{graphicx}
\usepackage{times}
\usepackage{graphics}
\usepackage{epsfig}
\usepackage{amsmath}

\begin{document}

\title{Comment on ``Passage Times for Unbiased Polymer Translocation through a Narrow Pore''}

\author{Kaifu Luo}
\affiliation{Laboratory of Physics, Helsinki University of Technology,
P.O. Box 1100, FIN-02015 TKK, Espoo, Finland}
\author{Tapio Ala-Nissila}
\affiliation{Laboratory of Physics, Helsinki University of Technology,
P.O. Box 1100, FIN-02015 TKK, Espoo, Finland}
\affiliation{Department of Physics, Box 1843, Brown University, Providence, Rhode Island 02912-1843, USA}
\author{See-Chen Ying}
\affiliation{Department of Physics, Box 1843, Brown University, Providence, Rhode Island 02912-1843, USA}
\author{Pawel Pomorski}
\affiliation{Department of Applied Mathematics, The University of Western Ontario, London, Ontario, Canada}
\author{Mikko Karttunen}
\affiliation{Department of Applied Mathematics, The University of Western Ontario, London, Ontario, Canada}

\date{August 21, 2007}
\pacs{87.15.Aa, 87.15.He, 36.20.-r}

\maketitle

One of the most fundamental quantities associated with polymer translocation through a nanopore is
the translocation time $\tau$ and its dependence on the chain length $N$.
In a recent Letter, Wolterink {\it et al.}~\cite{Wolterink06} present new results for unbiased
translocation using a 3D lattice model within the Monte Carlo (MC) simulation method. According to
their scaling argument $\tau \sim N^{1 + 2\nu}F(b/R_g)$, where $R_g$ is the radius of gyration, $b$
the pore width, the Flory exponent $\nu=0.75$ in 2D and 0.588 in 3D, and numerically the scaling
function $F(x) \sim x^{-0.38 \pm 0.08}$ for $x \rightarrow 0$.
This leads to $\tau \sim N^{2.40 \pm 0.08}$, which is in contradiction with the established
result by Chuang {\it et al.}~\cite{Chuang02} that $\tau \sim N^{2\nu+1}$, which gives
2.50 in 2D and 2.18 in 3D.

The original prediction that $\tau\sim N^{2\nu+1}$ \cite{Chuang02}
means that $\tau$ scales with $N$ in the same manner as the Rouse relaxation time of the
chain except for a larger prefactor, $\tau_R \sim R_g^2/D \sim N^{1 + 2\nu}$, where the diffusion
coefficient $D\sim 1/N$ within Rouse dynamics.
This result was recently corroborated by extensive numerical simulations based on the Fluctuating
Bond (FB)~\cite{Luo06} and Langevin Dynamics (LD) models with the bead-spring
approach~\cite{Huopaniemi06, Wei07}, where $\tau$ was found to scale as $N^{2.50 \pm 0.01}$ in 2D.

To resolve the apparent
discrepancy, we have
analyzed the approach of Wolterink 
{\it et al.}~\cite{Wolterink06} and performed additional high-accuracy numerical simulations using the FB
model with MC dynamics~\cite{Luo06} in 2D, and atomistic MD simulations using the
GROMACS~\cite{van-der-Spoel:05ws} simulation engine in 2D and 3D.
As in the MC and LD methods, explicit solvent hydrodynamics 
were excluded in our GROMACS simulations.
We find that within numerical accuracy, both the FB and MD
methods give the same scaling results in 2D, and thus here we present data
as obtained using GROMACS within the bead-spring model.
GROMACS is currently one of the most commonly used programs in soft matter
and biophysical simulations, and has also been used extensively by some of
us in various problems (see e.g., Ref.~\onlinecite{Patra:04po} and references therein).

\begin{figure}[tb]
\begin{center}
\includegraphics[width=6cm]{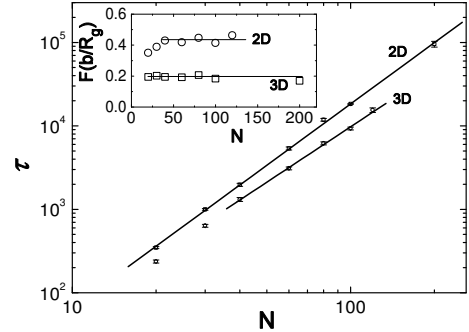}
\caption{
Scaling of escape times from GROMACS data. The slopes are
$2.44 \pm 0.03$ (2D)
 and $2.22 \pm 0.06$ (3D).
The insert shows the scaling function $F(b/R_g)$ ($\sim \tau/N^{1+2\nu}$) as a function of $N$.
}
\label{fig1}
\end{center}
\end{figure}

Our main results are summarized in Fig.~\ref{fig1}, where we find that
$\tau \sim N^{2.44 \pm 0.03}$ in 2D and $\tau \sim N^{2.22 \pm 0.06}$ in 3D in complete agreement
with Refs.~\onlinecite{Chuang02,Luo06,Huopaniemi06,Wei07}, ruling out results 
of Ref.~\onlinecite{Wolterink06} 
within the accuracy of the data.
In particular, we
find that the scaling function $F(b/R_g)$ becomes independent of $N$ for large $N$, 
as shown in the insert of Fig. 1. Thus, there is no discernible
correction to the exponent $1+2\nu$.

To verify our results independently, we also computed the 
squared change of the translocation coordinate $s(t)$, where we observed sub-diffusive behavior
$\langle (\Delta s(t))^2 \rangle \sim t^{\alpha}$, with $\alpha=0.807 \pm 0.002$ in 2D and
$\alpha=0.910 \pm 0.002$ in 3D. This again agrees with Chuang \textit{et al.}~\cite{Chuang02} who
obtained $\alpha = 2/(1+2 \nu)$, which gives $0.8$ in 2D and 0.92 in 3D.

Further theoretical support for the exponent $1+2\nu$ comes from two other independent
studies. In Ref.~\onlinecite{Luo06} it was analytically predicted and
also numerically confirmed that $\tau \sim (R_g+L)^2/D$ 
for a pore of length $L$, resulting from the fact that the mass center of the polymer 
moves a distance of $L$ ~\cite{Luo06}. 
For long pore $L \gg N$ we have $\tau \sim NL^2 \gg N^3$, which is longer than the reptation 
time of the chain $\sim N^3$.

In the second study a different approach was used: the starting 
point was driven translocation in which 
a pulling force $F$ is acting on one end of the chain.
In that case,
$\tau \sim N^{2\nu+1}$ can be analytically derived in the limit $F \rightarrow 0$
using well-established scaling functions for polymers under tension \cite{Huopaniemi07}.

To conclude, all the above \textit{independent} 
results, analytical and numerical,
and in particular the behavior of the scaling function $F(b/R_g)$,
confirm the result $\tau \sim N^{2\nu+1}$, and invalidate
the lattice model results of Ref.~\onlinecite{Wolterink06}.
The apparent discrepancy may be due to the artificial dynamics of the
lattice model of Ref.~\onlinecite{Wolterink06}, which
is based on the repton model for a single reptating polymer 
with the addition of sideways moves and reptation moves; 
although their lattice model works well for static
properties in bulk solutions
\cite{Heukelum07}, the dynamics in the presence of a narrow hole
may pose problems as the hole adds a
new length scale to the problem. An off-lattice approach,
such as used here is guaranteed to be free of any artifacts.

This work has been supported by the Academy of Finland through the
TransPoly and COMP CoE grants, and NSERC of Canada (M.\,K.).
We thank the SharcNet grid computing facility (www.sharcnet.ca)
for computer resources.

\end {document}